\documentclass[12pt,letterpaper]{article}
\usepackage[utf8]{inputenc}
\usepackage{tabularx}
\usepackage{array}
\usepackage{placeins}
\usepackage{longtable}
\usepackage{newtxtext,newtxmath}
\usepackage{lipsum}
\usepackage{amsmath}
\usepackage{amsfonts}
\usepackage{graphicx}
\usepackage{cite}
\usepackage{ulem}
\usepackage{caption}
\usepackage[top=1in,bottom=1in,left=1in,right=1in]{geometry} 
\usepackage{float}
\usepackage{hyperref}
\usepackage{booktabs}
\usepackage[export]{adjustbox}
\usepackage{float}
\usepackage{multirow}
\usepackage{pifont}
\usepackage[ ]{authblk}
\usepackage{chemarrow}
\usepackage{comment}
\usepackage{makecell}
\usepackage{soul}
\UseRawInputEncoding
\hypersetup{
    colorlinks=true,
    linkcolor=blue,
    filecolor=magenta,      
    urlcolor=blue,
    pdftitle={Overleaf Example},
    pdfpagemode=FullScreen,
}
\usepackage{multicol}
\begin{document}

%
%
\title{
Modeling and Experimental Validation of Bubble-Induced Performance Losses in Microstructured Copper Electrodes for Alkaline Water Electrolysis}
%
%

%
%
\author[1,$\dag$]{Auronno Ovid Hussain}
\author[1,$\dag$]{Adith Saha Dipro}
\author[1,*]{Muhammad Anisuzzaman Talukder}
\affil[1]{\small{Department of Electrical and Electronic Engineering \\

Bangladesh University of Engineering and Technology, Dhaka 1205, Bangladesh}} 

\affil[$\dag$]{\small{\it{These authors contributed equally to this work.}}}
\affil[*]{\small{\it{anis@eee.buet.ac.bd}}}
%
%

\date{ }
\maketitle
\sloppy

\begin{abstract}
Water electrolysis is a promising technology for sustainable hydrogen (H$_2$) production. However, gas bubble formation significantly degrades the electrochemical surface area ($A_{\rm EC}$) and the electrocatalytic activity of microstructured electrodes in water electrolyzer cells. Existing theoretical models fail to connect computational fluid dynamics (CFD) and electrochemistry to analyze how bubble dynamics affect the performance of gas-evolving microstructured electrodes. This work develops a multiphysics-based theoretical model that couples two-phase CFD with Butler-Volmer and Faradaic kinetics, incorporating exchange current density, cell current, gravitational acceleration, surface tension, and fluid properties such as density and viscosity. The model quantifies essential electrochemical parameters, such as the enhancement in $A_{\rm EC}$ and the overpotential ($\eta$), which deviate by less than 7\% and 15\%, respectively, from the experimental results. Evaluations of the fabricated microstructured electrodes using polarization curves, cyclic voltammetry, electrochemical impedance spectroscopy, and linear polarization resistance techniques demonstrate reduced $\eta$ by 2.49--21.39\% and 9.55--24.32\% for oxygen (O$_2$) and H$_2$ evolution, respectively. Furthermore, the model reveals the interplay between enhanced geometric surface area and bubble-transport limitations, facilitating the optimization and analysis of structural parameters. These findings establish a physics-informed roadmap for designing optimal, robust, and industry-scalable microstructured electrodes for efficient H$_2$ generation via water electrolysis.
\end {abstract}

\section{Introduction}
The global energy demand is increasing due to a growing world population, rapid industrialization, and rising energy consumption driven by advancing technology \cite{sun2025lignin}. Our efforts to overcome this obstacle have introduced global climate change, which is one of the most crucial challenges of the 21$^{\rm st}$ century \cite{ren2024recent, mamun2024techno}. Growing concerns over global climate change have motivated scientists and engineers to investigate sources of greener, cleaner energy fuels for the future \cite{zheng2022future}. Hydrogen (H$_2$) gas has a promising potential as an alternative to fossil fuels due to its high energy content and zero carbon emissions \cite{zhang2023electrochemical, mamun2025re}. Although various technologies produce H$_2$ fuel through different chemical processes and energy sources, water electrolysis is considered one of the most promising techniques due to its decarbonized approach to H$_2$ generation and utilization \cite{wen2026electrolyzing}. Water electrolysis also offers several advantages, including the production of high-purity H$_2$, its non-toxic nature, and the ability to use a wide range of power sources \cite{el2023hydrogen}. The main components of water electrolyzer cells include electrodes, catalysts, ion exchange membranes, electrolytes, and stack modules \cite{du2022anion}. Among these, the electrode is the most critical part of the electrolyzer cell, as it facilitates charge transfer to the electrolyte, producing H$_2$ and oxygen (O$_2$) gases through the hydrogen evolution reaction (HER) and the oxygen evolution reaction (OER), respectively \cite{xie2022oxygen, zhang2023superaerophilic}.

Noble metals, such as platinum (Pt), iridium (Ir), ruthenium (Ru), rhodium (Rh), and palladium (Pd), are excellent candidates for electrode materials in water electrolysis due to their high electrocatalytic activity and corrosion resistance \cite{mamun2024enhancing}. However, the use of these metals is limited because of their low natural abundance and high cost \cite{liu2022recent}. As a result, non-noble metals, including copper (Cu), cobalt (Co), and nickel (Ni), have shown promise as efficient electrodes for water electrolyzers due to their natural abundance, cost-effectiveness, and tunable electronic structures \cite{wang2026recent}. Due to their slower reaction kinetics and higher corrosion rates, many techniques, such as alloying, doping, and structuring, have been used to enhance their electrocatalytic activity and reduce their corrosion rates \cite{mamun2025advancing, wu2025corrosion}. Despite the implementation of various techniques, micro and nanostructuring have emerged as promising strategies because they increase the number of active sites on the electrode surface, enhance the electrochemical surface area ($A_{\rm EC}$), and facilitate rapid charge transfer \cite{chandrasekaran2022developments}. The ease of fabrication, scalability, and high electrochemical stability have made microstructures attractive for industrial water electrolyzers \cite{du2024continuous}.  

Despite much effort in the surface engineering of non-noble metals, bubble dynamics remains a crucial aspect of efficient electrode design for water electrolysis \cite{riaz2025water}. Microstructures fabricated on the electrode surface facilitate gas bubble formation and promote bubble pinning by lowering the nucleation energy barrier \cite{xie2026recent, zhao2023electrolyte}. Due to bubble pinning and surface coverage effects, the area inside these microstructures becomes inaccessible for gas evolution reactions, thereby reducing $A_{\rm EC}$. Additionally, the pinned bubbles reduce the local electrolyte conductivity, thereby increasing Ohmic losses. 

Recently, the investigation into the bubble dynamics in water electrolyzers has been carried out using analytical models, microfluidics, finite element method (FEM), and artificial intelligence (AI)-based optimization \cite{lu2026three, li2026euler, sarma2023computational, li2021bubbles, ross2025impact}. Lu et al.~developed a three-dimensional (3D) model of an alkaline water electrolyzer in COMSOL Multiphysics, coupling electrochemical reactions with fluid flow and mass transfer \cite{lu2026three}. However, this model cannot account for microstructures on the surface of pristine metal electrodes. An Euler-Lagrange (E-L) model that accounts for turbulent flow was developed to investigate the behavior of H$_2$ gas bubbles in alkaline water electrolyzers \cite{li2026euler}. Since the E-L model solves for a two-phase fluid system, it cannot accurately distinguish between the electrolyte and gas bubbles. Later, Sarma et al.~employed the phase-field method to develop a more robust and advanced model for analyzing the movement of gas bubbles in hematite nanostructured electrodes \cite{sarma2023computational}. Apart from this, Li et al.~developed a microfluidics-based analytical model, which measures the bubble growth rate and detachment kinetics \cite{li2021bubbles}. Additionally, the detrimental effect of gas bubbles on cell overpotential loss, assuming a bare electrode surface, has been modeled by Ross et al. \cite{ross2025impact}. None of these existing models can be used to analyze the electrocatalytic activity or overpotential losses of microstructured electrodes in the presence of gas bubbles.

In this work, we have developed a multiphysics theoretical model by integrating two-phase computational fluid dynamics (CFD) with electrochemical theories, including Butler-Volmer and Faradaic kinetics, to examine the effect of bubbles on the electrochemical performance of microstructured electrodes in water electrolyzers. This model, implemented using FEM, quantifies the enhancement in electrochemical surface area ($E_{\rm A_{EC}}$) and the overpotential ($\eta$) of microstructured electrodes used in water electrolysis. The model incorporates the exchange current density ($J_0$), cell current ($J_{\rm cell}$), surface coverage coefficient ($\theta$), surface tension ($\mathbf{F_{\rm st}}$), gravitational acceleration ($\mathbf{g}$), and fluid properties, including viscosity ($\mu$) and density ($\rho$). 

We have fabricated three microstructured copper (Cu) electrodes with holes of varying diameters ($D$) and numbers to validate the developed model. The electrochemical performance of these electrodes was analyzed using polarization curves (PC), cyclic voltammetry (CV), electrochemical impedance spectroscopy (EIS), and linear polarization resistance (LPR) measurements. PC measurements show that the electrodes reduce $\eta$ for O$_2$ and H$_2$ evolution by 2.49--21.39\% and 9.55--24.32\%, respectively, relative to bare Cu, indicating enhanced OER and HER kinetics. Additionally, the EIS measurements reveal a 2.56--17.75\% reduction in the charge-transfer resistance ($R_{\rm ct}$) relative to bare Cu, which demonstrates enhanced interfacial charge transfer. 

The $E_{\rm A_{EC}}$ and $\eta$ values extracted from experimental results remain within $7\%$ and 15\% of the model predictions, respectively, thereby confirming the model's accuracy and applicability across different structural parameters. Additionally, the model captures the trade-off between geometric scaling effects and bubble-induced transport limitations, providing a physics-based framework for optimizing $D$. Furthermore, the model demonstrates the effect of interhole separation ($a$) on electrocatalytic activity. Therefore, the developed model serves as a predictive design framework for energy-efficient microstructured electrodes in water electrolyzers.

\section{Multiphysics CFD-Electrochemical Model for Bubble Dynamics in Microstructured Electrodes}

\subsection{Model Development}
A water electrolyzer cell contains two different types of fluids \cite{el2023hydrogen}. One is the electrolyte solution, while the other is the gas bubbles generated under operating conditions. Therefore, we have considered the electrolyzer as a two-phase fluid system. To describe the dynamics of this system, the incompressible Navier-Stokes equations have been used \cite{al2023dynamics}. The momentum equation is given as follows \cite{al2023dynamics}
\begin{equation}
    \rho \left( \frac{\partial \mathbf{u}}{\partial t} + \mathbf{u} \cdot \nabla \mathbf{u} \right)
    = - \nabla p + \nabla \cdot \mu \left( \nabla \mathbf{u} + \nabla \mathbf{u}^{\rm T} \right) + \rho \mathbf{g} + \mathbf{{F_{\rm st}}},
    \label{eq:1}
\end{equation}
where $\mathbf{u}$ and $p$ represent two-phase fluid velocity and pressure, respectively. The continuity equation is given as follows \cite{al2023dynamics}
\begin{equation}
    \nabla \cdot \mathbf{u} = 0.
    \label{eq:2}
\end{equation}
Equations \eqref{eq:1} and \eqref{eq:2} together describe the two-phase flow within the electrochemical system. Since the Navier-Stokes equations alone cannot distinguish between the liquid and gaseous phases, we have used the level set transport equation to determine the motion of the bubbles. The transport equation is given by \cite{al2023dynamics} 
\begin{equation}
    \frac{\partial \phi}{\partial t} + \mathbf{u} \cdot \nabla \phi = \gamma \nabla \cdot \left( \epsilon \nabla \phi - \phi \left( 1 - \phi \right) \frac{\nabla \phi}{\left| \nabla \phi \right|} \right),
    \label{eq:3}
\end{equation}
where $\gamma$ is the initialization parameter, $\phi$ is the level set variable, and $\epsilon$ is the thickness of the diffuse fluid interface. The parameter $\phi$ ranges from 0 to 1, where $\phi=0$ and $\phi=1$ represent two distinct fluid phases. The parameter $\epsilon$ controls the thickness of the diffuse interface over which $\phi$ transitions smoothly from zero to 1 rather than changing discontinuously and is chosen as \cite{al2023dynamics}
\begin{equation}
    \epsilon = \frac{h_c}{2},
    \label{eq:4}
\end{equation}
where $h_c$ is the largest mesh element size at the interface of the two fluid phases. Using $\phi$, we can define $\rho$ and $\mu$ as follows \cite{al2023dynamics}
\begin{subequations}
\begin{equation}
    \rho = \rho_{\rm liq} + \left(\rho_{\rm gas} - \rho_{\rm liq} \right)\phi,
    \label{eq:5a}
\end{equation}
\begin{equation}
    \mu = \mu_{\rm liq} + \left(\mu_{\rm gas} - \mu_{\rm liq} \right)\phi,
    \label{eq:5b}
\end{equation}
\end{subequations}
where the subscripts ``liq'' and ``gas'' denote the electrolyte and the gas bubbles, respectively. Equations \eqref{eq:1}--\eqref{eq:3} are the governing equations for our model.

The gas bubbles generated during water electrolysis reduce the electrocatalytic activity of the electrode surface \cite{wu2023bubbles}. Conversely, a higher gas removal rate enhances surface reactivity \cite{sarma2023computational}. The mass of the generated gas during electrolysis, $m_{\rm gen}$, can be measured using Faraday's law of electrolysis, which is given as follows \cite{ghaib2018power}
\begin{equation}
    m_{\rm gen} = \frac{J_{\rm cell}M_{\rm gas}A_{\rm geo}t}{zF},
    \label{eq:6}
\end{equation}
where $M_{\rm gas}$ is the molar mass of the gas, $A_{\rm geo}$ is the geometric surface area, $t$ is the duration of electrolysis, $z$ is the number of electrons transferred, and $F$ is the Faraday constant. We modify Eq.~\eqref{eq:6} to obtain the expression for the gas generation rate, $V_{\rm gen}$, and it is given as follows \cite{bard2022electrochemical}
\begin{equation}
    V_{\rm gen} = \frac{J_{\rm cell}M_{\rm gas}A_{\rm geo}}{zF\rho_{\rm gas}}.
    \label{eq:7}
\end{equation}

The gas removal rate, $Q$, can be expressed by \cite{cengel2013ebook}
\begin{equation}
    Q = A_{\rm rem}\phi u,
    \label{eq:8}
\end{equation}
where $A_{\rm rem}$ is the surface area through which gas bubbles are escaping. The effect of $V_{\rm gen}$ and $Q$ on $\theta$ is given by \cite{bard2022electrochemical, sarma2023computational}
\begin{equation}
    \theta \propto \frac{V_{\rm gen}}{Q}. 
    \label{eq:9}
\end{equation}
Electrochemical theories suggest that as $V_{\rm gen} \rightarrow 0$, $\theta \rightarrow 0$. Additionally, as $Q \rightarrow 0$, $\theta \rightarrow 1$. In accordance with these limiting cases, we propose a simple expression for $\theta$, which is given as follows
\begin{equation}
    \theta = \frac{V_{\rm gen}}{V_{\rm gen} + Q}.
    \label{eq:10}
\end{equation}
The parameter $A_{\rm EC}$ of the microstructured electrode can be determined using \cite{ross2025impact}
\begin{equation}
    A_{\rm EC} = \left(1-\theta\right) A_{\rm {geo}}.
    \label{eq:11}
\end{equation}
Equation \eqref{eq:11} illustrates that $A_{\rm EC}$ and $A_{\rm geo}$ differ by $\left(1-\theta\right)$. Only in the idealized case of $\theta = 0$, these two areas are equal. 

The parameter $\eta$ is given as follows \cite{jang2021numerical}
\begin{equation}
    \eta = \frac{RT}{\alpha zF} \log\left(\frac{J_{\rm cell}}{J_{\rm 0,eff}}\right),
    \label{eq:12}
\end{equation}
where $R$ is the universal molar gas constant, $T$ is the system temperature, $\alpha$ is the reaction transfer coefficient, and $J_{\rm 0,eff}$ is the effective exchange current density. The parameter $J_0$ is related to $J_{\rm 0,eff}$ by the following expression \cite{jang2021numerical}
\begin{equation}
    J_{\rm 0,eff} = \left(1-\theta\right) J_0.
    \label{eq:13}
\end{equation}
By combining Eqs.~\eqref{eq:12} and \eqref{eq:13}, we obtain 
\begin{equation}
    \eta = \frac{RT}{\alpha zF} \log \left [\frac{J_{\rm cell}}{\left(1-\theta\right)J_0}\right].
    \label{eq:14}
\end{equation}
Equation \eqref{eq:14} is further modified by using Eq.~\eqref{eq:10}, and the modified expression is given as follows
\begin{equation}
    \eta = \frac{RT}{\alpha zF} \log \left [\frac{J_{\rm cell}}{\left(1-\frac{V_{\rm gen}}{V_{\rm gen} + Q}\right)J_0}\right].
    \label{eq:15}
\end{equation}
We simplify Eq.~\eqref{eq:15}, and the modified expression is given by
\begin{equation}
    \eta = \frac{RT}{\alpha zF} \log \left(\frac{J_{\rm cell}}{J_0}\right) + \frac{RT}{\alpha zF} \log \left(1 + \frac{V_{\rm gen}}{Q}\right).
    \label{eq:16}
\end{equation}
Equation \eqref{eq:16} illustrates that $\eta$ has two components. The first one is the activation overpotential loss, $\eta_{\rm act}$, while we denote the other one as the transport overpotential loss, $\eta_{\rm tr}$, which is related to the motion of gas bubbles in the electrolyzer cell \cite{mamun2023effects}.

\subsection{Simulation Model}

\begin{figure}[ht]
    \centering
    \includegraphics[width =\linewidth]{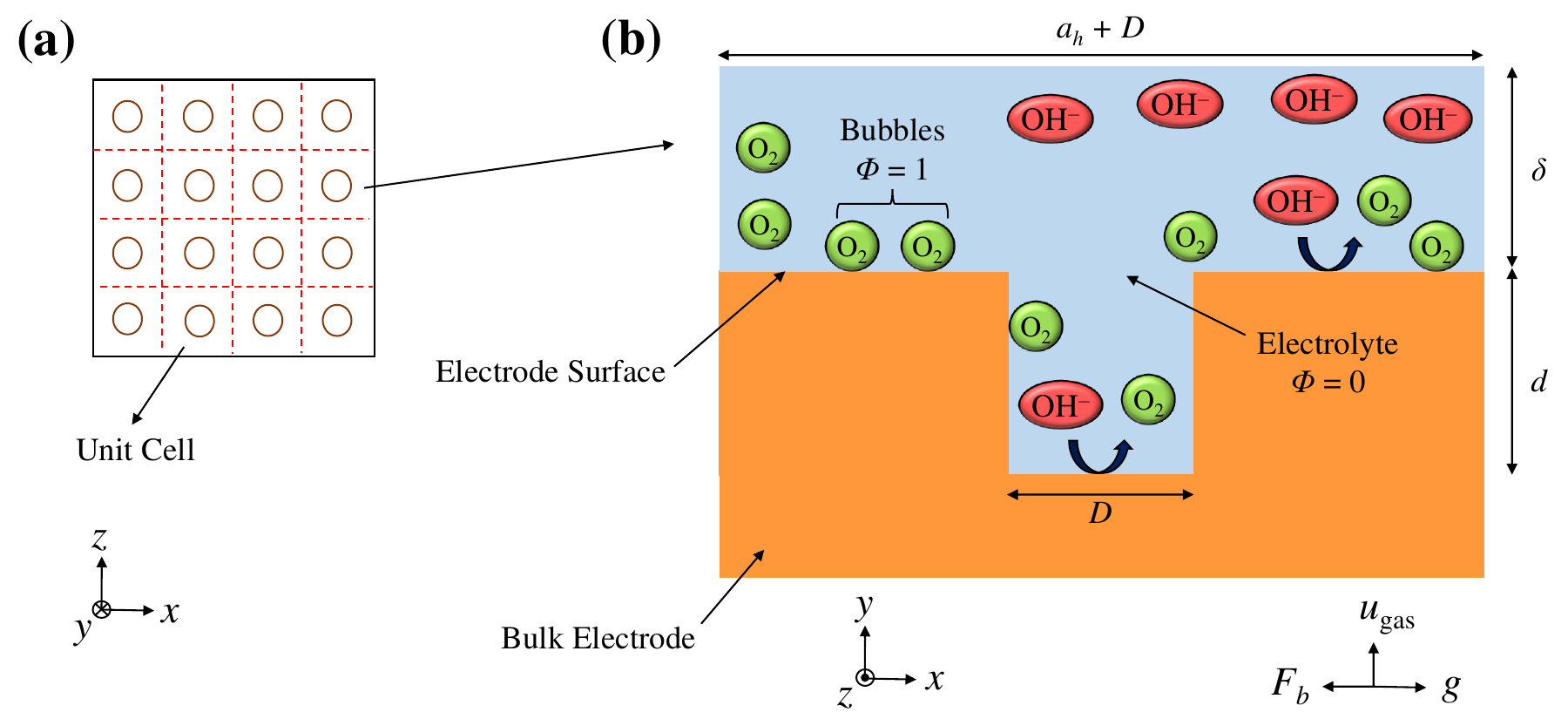}
    \caption{Two-dimensional (2D) schematic illustrations of (a) the fabricated electrode surface, which is divided into several unit cells, and (b) the vertical cross-section of a unit cell and its surroundings. The parameter $\mathbf{u_{\rm gas}}$ is the velocity of the gas bubbles at the electrode surface, $\mathbf{g}$ is the gravitational acceleration, $\mathbf{F_b}$ is the buoyancy force, and $\delta$ is the diffusion layer thickness, $a_h$ is the horizontal separation between microholes, and $D$ and $d$ are the diameter and depth of a microhole. Here, we assume the electrode is an anode in an alkaline water electrolyzer cell.}
    \label{Fig 1}
\end{figure}

The laminar flow and level set study modules in the COMSOL Multiphysics software package were used to solve the developed model discussed in Sec.~2.1 and investigate the bubble dynamics and electrochemical performance of microstructured electrodes. Figure \ref{Fig 1}(a) illustrates the two-dimensional (2D) schematic illustration of the fabricated electrode surface, and Fig.~\ref{Fig 1}(b) represents the vertical cross-section of the electrode structure. Figure \ref{Fig 1}(b) also highlights the essential physical phenomena governing the bubble production and detachment. We have considered a planar electrode structure with microholes on each surface. We assumed that the electrode consists of unit cells, each containing a microhole and a flat surface, for computational purposes. The unit cell repeats periodically in the in-plane $x$ and $z$ directions of the electrode surface. The periodicities were determined from the corresponding interhole separations in the $x$ ($a_h$) and $z$ ($a_v$) directions.

\begin{figure}
    \centering
    \includegraphics[width =\linewidth]{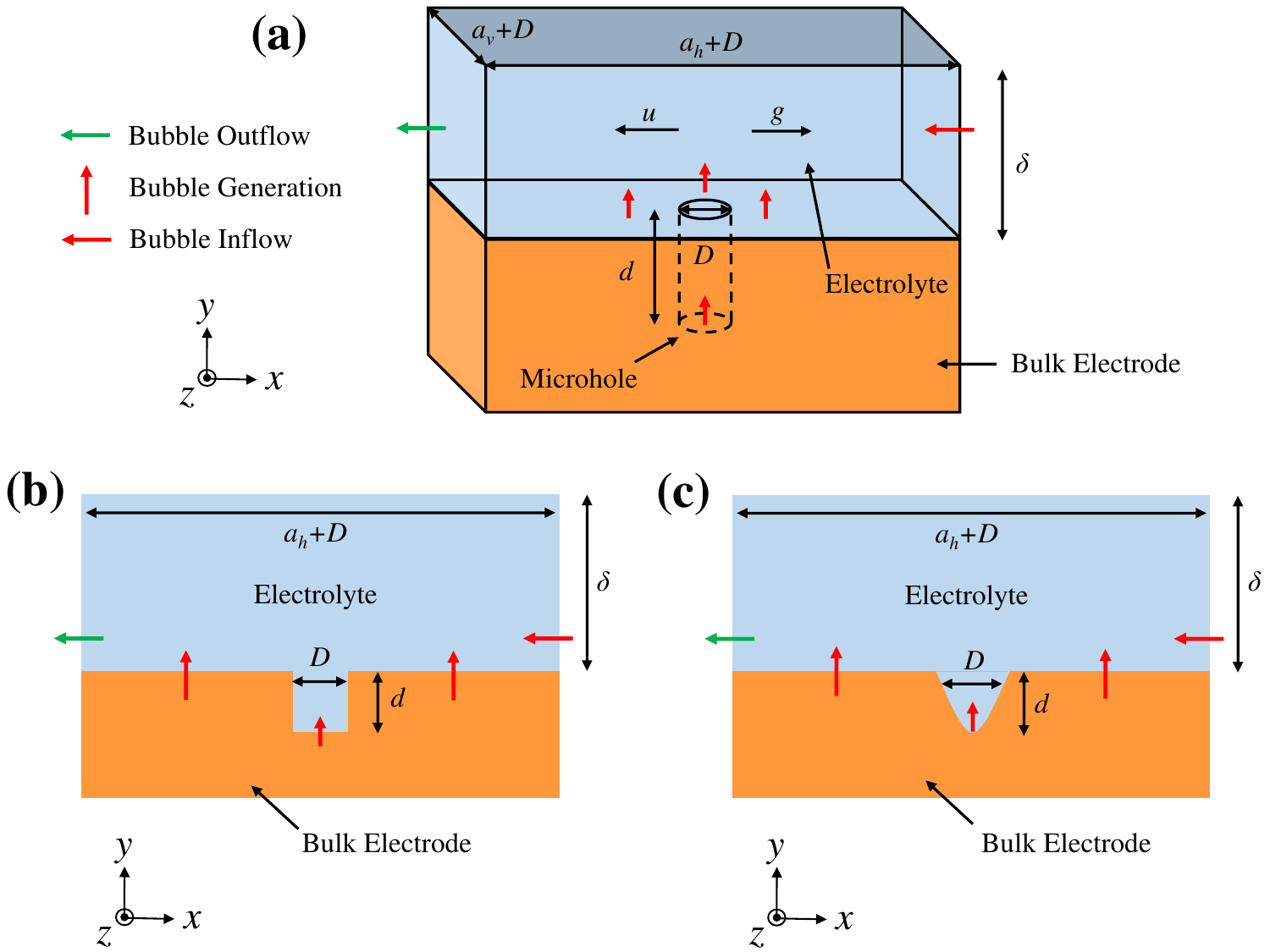}
    \caption{(a) 3D simulation structure for the model with both electrode surface and electrolyte. Vertical cross-sections of the (b) idealized and (c) realistic 3D simulation structure. The parameters $a_h$ and $a_v$ are the horizontal and vertical separations between microholes, respectively, while $D$ and $d$ are the diameter and depth of the microholes, respectively. The parameters $u$, $g$, and $\delta$ denote the two-phase fluid velocity, the gravitational acceleration, and the diffusion layer thickness, respectively.}
    \label{Fig 2}
\end{figure}

Figure \ref{Fig 2} illustrates the various structures simulated using the developed model. Figure \ref{Fig 2}(a) presents the 3D structure used in the simulation. Due to computational cost, we have also worked with 2D cross-sections of the 3D structure, as shown in Figs.~\ref{Fig 2}(b) and \ref{Fig 2}(c). It is worth mentioning that the microhole is modeled as a rectangle in Fig.~\ref{Fig 2}(b). However, due to the nature of the fabrication technique, the mouth and the bottom portions of the hole are expected to be wider and narrower, respectively \cite{diver2004micro, mincuzzi2019novel}. Therefore, we have used curvature-induced holes in the 2D structure, as shown in Fig.~\ref{Fig 2}(c). For the remainder of this paper, the structures shown in Fig.~\ref{Fig 2} will be referred to as 3D, 2D, and 2D (Curved), respectively. The key parameters used to simulate these structures are summarized in Table \ref{Table 1}.

In our model, bubbles appear within a unit cell through two distinct mechanisms. A portion of the gas bubbles is generated by electrochemical reactions at the electrode surface, while others enter the simulated unit cell from the nearby unit cell. Due to the buoyancy force, $F_b$, the bubbles will move towards the left. We define $\phi = 1$ and  $\phi = 0$ to specify the bubbles and electrolyte, respectively. 

Appropriate initial and boundary conditions were chosen to solve Eqs.~\eqref{eq:1}--\eqref{eq:3} for the microstructured electrodes. Initially, no electrochemical reactions occur on the electrode surface, indicating no gas generation in the electrolyzer \cite{el2023hydrogen}. Therefore, we have assumed $\phi=0$, which represents the electrolyte phase, throughout the system. Additionally, we have considered $u=0$ as the initial condition, since the absence of gas evolution implies no induced fluid motion. Furthermore, $p$ has been initialized to zero because there is no pressure buildup due to gas generation. Here, $p$ denotes gauge pressure, the difference between the actual pressure and atmospheric pressure. The parameter $p$ is set to zero at the topmost boundary of the unit cell shown in Fig.~\ref{Fig 1}(b) since it is open to the bulk electrolyte. The velocity of the bubbles at the electrode surface, $u_{\rm gas}$, was calculated using
\begin{equation}
    u_{\rm gas} = \frac{J_{\rm cell}M_{\rm gas}\left(1-\phi\right)}{zF\rho_{\rm gas}}.
    \label{eq:17}
\end{equation}

\begin{longtable}{ll}
\caption{Key parameters of the multiphysics CFD-electrochemical model. \label{Table 1}} \\
\hline \hline
Parameter & Value \\ \hline
Universal molar gas constant, $R$ & 8.314 JK$^{-1}$mol$^{-1}$ \\
System temperature, $T$ & $298$ K \\
Number of electrons transferred, $z$ & $4$ \\
Faraday constant, $F$ & $96485$ Cmol$^{-1}$ \\
Exchange current density, $J_0$ & $5.524 \times10^{-4}$ Am$^{-2}$ \\
Density of KOH, $\rho_{\rm liq}$ & $1045$ kgm$^{-3}$ \\
Viscosity of KOH, $\mu_{\rm liq}$ & $9.3\times10^{-4}$ Pa.s \\
Density of O$_2$, $\rho_{\rm gas}$ & $1.22$ kgm$^{-3}$ \\
Viscosity of O$_2$, $\mu_{\rm gas}$ & $1.8\times10^{-5}$ Pa.s \\
Molar mass of O$_2$, $M_{\rm gas}$ & $0.032$ kgmol$^{-1}$ \\
Gravitational acceleration, $g$ & $9.81$ ms$^{-2}$ \\
Surface tension coefficient, $\sigma$ & $0.073$ Nm$^{-1}$ \\
Reinitialization parameter, $\gamma$ & $0.005$ ms$^{-1}$ \\
\hline \hline
\end{longtable}

The 3D unit cell of the microstructured electrode, shown in Fig.~\ref{Fig 2}(a), considers an idealized cylindrical microhole. Therefore, the $A_{\rm geo}$ of the unit cell is given as follows
\begin{equation}
    A_{\rm geo} = 2\pi rd + \left(a_h+2r\right)\left(a_v+2r\right),
    \label{eq:18}
\end{equation}
where $r$ and $d$ are the radius and depth of the microhole. The dimensions $a_h$ and $a_v$ are given by the following expressions   
\begin{subequations}
\begin{equation}
    a_h = \frac{L-2rn_h}{n_h+1},
    \label{eq:19a}
\end{equation}
\begin{equation}
    a_v = \frac{L-2rn_v}{n_v+1},
    \label{eq:19b}
\end{equation}
\end{subequations}
where $L$ is the length of the electrode surface, and $n_h$ and $n_v$ are the total number of microholes in each row and column on the surface, respectively. Additionally, $A_{\rm rem}$ is expressed as follows
\begin{equation}
    A_{\rm rem} = \left(a_v+2r\right)\delta,
    \label{eq:20}
\end{equation}
where $\delta$ is the thickness of the diffusion layer.

$E_{\rm A_{\rm EC}}$ and the enhancement in $A_{\rm geo}$, $E_{\rm A_{\rm geo}}$, are defined using the following expressions
\begin{subequations}
\begin{equation}
   E_{\rm A_{\rm geo}} = \frac{A_{\rm geo,s}-A_{\rm geo,b}}{A_{\rm geo,b}} \times 100\%,
   \label{eq:21a}
\end{equation}
\begin{equation}
   E_{\rm A_{\rm EC}} = \frac{A_{\rm EC,s}-A_{\rm EC,b}}{A_{\rm EC,b}} \times 100\%,
   \label{eq:21b}
\end{equation}
\end{subequations}
where the subscripts ``s'' and ``b'' stand for microstructured and bare electrodes, respectively. $E_{\rm A_{\rm EC}}>0$ is desirable as it indicates that the enhancement in electrocatalytic activity due to microstructuring exceeds the losses caused by bubble coverage. However,  $E_{\rm A_{\rm EC}}<0$ is undesirable because it demonstrates that the bubble-induced surface blockage dominates, resulting in a net reduction in electrocatalytic activity. 

\subsection{Conditions for Designing Optimal Microstructured Electrodes}
The parameter $E_{\rm A_{EC}}$, given by Eq.~\eqref{eq:21b}, represents the increase in catalytic sites compared to the pristine electrode. There are two competing physical mechanisms that govern the enhancement in electrocatalytic activity due to microstructuring. One is the geometric scaling, which increases the number of active sites on the surface. The other one is the bubble detachment kinetics, which becomes more prominent at high $J_{\rm cell}$ due to faster gas evolution. Given the presence of these competing processes, optimal operating conditions must be met. Therefore, we combine Eqs.~\eqref{eq:10} and \eqref{eq:11} with Eq.~\eqref{eq:21b}, which yields the modified expression for $E_{\rm A_{EC}}$ as follows
\begin{equation}
    E_{\rm A_{EC}} = \left(\frac{Q_s}{Q_b}\right) \left(\frac{V_{\rm gen}+Q_b}{V_{\rm gen}+Q_s}\right) \left(\frac{A_{\rm geo,s}}{A_{\rm geo,b}}\right) - 1.
    \label{eq:22}
\end{equation}
Equation \eqref{eq:22} shows that $E_{\rm A_{EC}}$ is influenced by geometric scaling, bubble detachment, and bubble generation. We explore how these dependencies change at two terminal cases. At low $J_{\rm cell}$, $V_{\rm gen}\rightarrow0$, and Eq.~\eqref{eq:22} is modified to
\begin{equation}
    E_{\rm A_{EC}} \approx  \left(\frac{A_{\rm geo,s}}{A_{\rm geo,b}}\right) - 1.
    \label{eq:23}
\end{equation}
Equation \eqref{eq:23} denotes that during negligible bubble production, $E_{\rm A_{EC}}$ solely depends on the geometric scaling due to microstructuring. Additionally, we modify Eq.~\eqref{eq:22} for high $J_{\rm cell}$ applications, which is given as follows
\begin{equation}
    E_{\rm A_{EC}} \approx \left(\frac{Q_s}{Q_b}\right) \left(\frac{A_{\rm geo,s}}{A_{\rm geo,b}}\right) - 1.
    \label{eq:24}
\end{equation}
Equation \eqref{eq:24} demonstrates that under industrial operating conditions, the enhancement in electrochemical activity becomes increasingly governed by bubble removal kinetics rather than geometric surface area alone \cite{jang2021numerical}. Therefore, microstructures with poor bubble-removal characteristics may exhibit reduced catalytic activity despite having larger geometric surface areas.

\section{Experimental Procedure}

\subsection{Material Preparation}
\begin{figure}
    \centering
    \includegraphics[width =\linewidth]{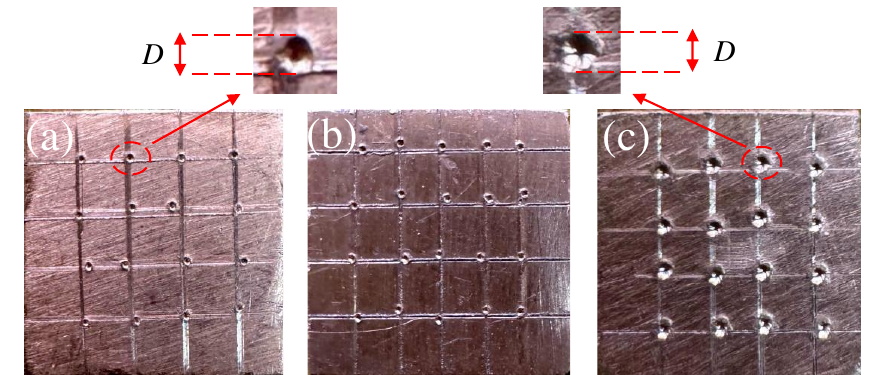}
    \caption{Images of microstructured electrodes (a) S1 ($D = 112.1$ $\mu$m, $d = 179.894$ $\mu$m, $16$ holes per surface), (b) S2 ($D = 90.4$ $\mu$m, $d = 95.3$ $\mu$m, $20$ holes per surface), and (c) S3 ($D = 506.1$ $\mu$m, $d = 161.36$ $\mu$m, $16$ holes per surface), where $D$ and $d$ denote the diameter and depth of the microhole.}
    \label{Fig 3}
\end{figure}

Industry-grade Cu sheets with a purity of 99.90\% were cut into $10 \times 10 \times 1$ mm-sized plates. The microholes were fabricated on the Cu samples using selective laser melting (SLM), as shown in Fig.~\ref{Fig 3}. The microstructured electrodes are named as S1 ($D = 112.1$ $\mu$m, $d = 179.894$ $\mu$m, $16$ holes per surface), S2 ($D = 90.4$ $\mu$m, $d = 95.3$ $\mu$m, $20$ holes per surface), and S3 ($D = 506.1$ $\mu$m, $d = 161.36$ $\mu$m, $16$ holes per surface), where $d$ denotes the depth of the microhole. The full distribution of hole sizes and depths is shown in Table \ref{Table 2}. We have measured $D$ and $d$ for all microholes and computed their ranges and means. 

To prepare the electrode surfaces for electrochemical analysis, the plates were cleaned and rinsed with acetone, ethanol, and deionized (DI) water for 5 minutes each. Additionally, the plates were carefully polished sequentially with emery paper having grits of 600, 800, 1000, 1200, and 2400 before each test. The DI water, sourced from the German Lab, had a conductivity of $<5.0$ $\mu$Scm$^{-1}$, and a pH of 8. Acetone and ethanol were obtained from Sigma-Aldrich, and anhydrous potassium hydroxide (KOH) pellets from Merck for preparing alkaline solutions. All reagents used in this work had a purity of $\geq99.5$\%.
\begin{longtable}{llcccc}
\caption{Parameters of microstructured electrodes. \label{Table 2}} \\
\hline \hline
Electrode & No.~of Holes & \multicolumn{2}{c}{$D$ ($\mu$m)} & \multicolumn{2}{c}{$d$ ($\mu$m)}\\
Sample & / Surface & Range & Mean & Range & Mean \\
\hline
S1 & $16$ & $76.1-144.3$ & $112.1$ & $130-240$ & $179.894$ \\
S2 & $20$ & $62.5-122.4$ & $90.4$ & $50-190$ & $95.3$ \\
S3 & $16$ & $430.8-634.2$ & $506.1$ & $120-240$ & $161.36$ \\
\hline
\hline
\end{longtable}

\subsection{Electrochemical Measurements}
The experiment was carried out in $250$ mL sealed H-cells using a standard three-electrode method. Bare Cu plate and structured samples, such as S1, S2, and S3, were used as the working electrode (WE). Pt plate and saturated Ag/AgCl were used as the counter electrode (CE) and the reference electrode (RE), respectively. The interelectrode separation was 1 cm. The concentration of the alkaline electrolyte was $1$ M, and the electrolyte was kept stagnant to ensure that the mass transfer occurred primarily by diffusion. The temperature was kept constant at 25 $^{\circ}$C. The room pressure was $1$ atm throughout the experiment.  

A Corrtest CS350M electrochemical workstation was used to perform electrochemical analyses, including PC, CV, EIS, and LPR tests. The electrochemical measurements were conducted after the open-circuit potential (OCP) had stabilized. The PC for OER and HER were obtained by varying the applied voltage, $E$, from $+0.4$ to $+0.9$ V and $-1.7$ to $-1.3$ V with respect to RE, respectively. Additionally, CV was carried out by varying the scan rate, $v$, from $10$ to $60$ mVs$^{-1}$, while the LPR analyses were carried out at a $v$ of 10 mVs$^{-1}$. Furthermore, EIS was performed at the OCP over the frequency range from 0.01 Hz to 100 kHz, with $10$ points per decade. The AC amplitude of the sinusoidal wave was 10 mV.

\section{Results and Discussion}

\subsection{Bifunctional Characteristics}
\begin{figure}[ht]
    \centering
    \includegraphics[width =\linewidth]{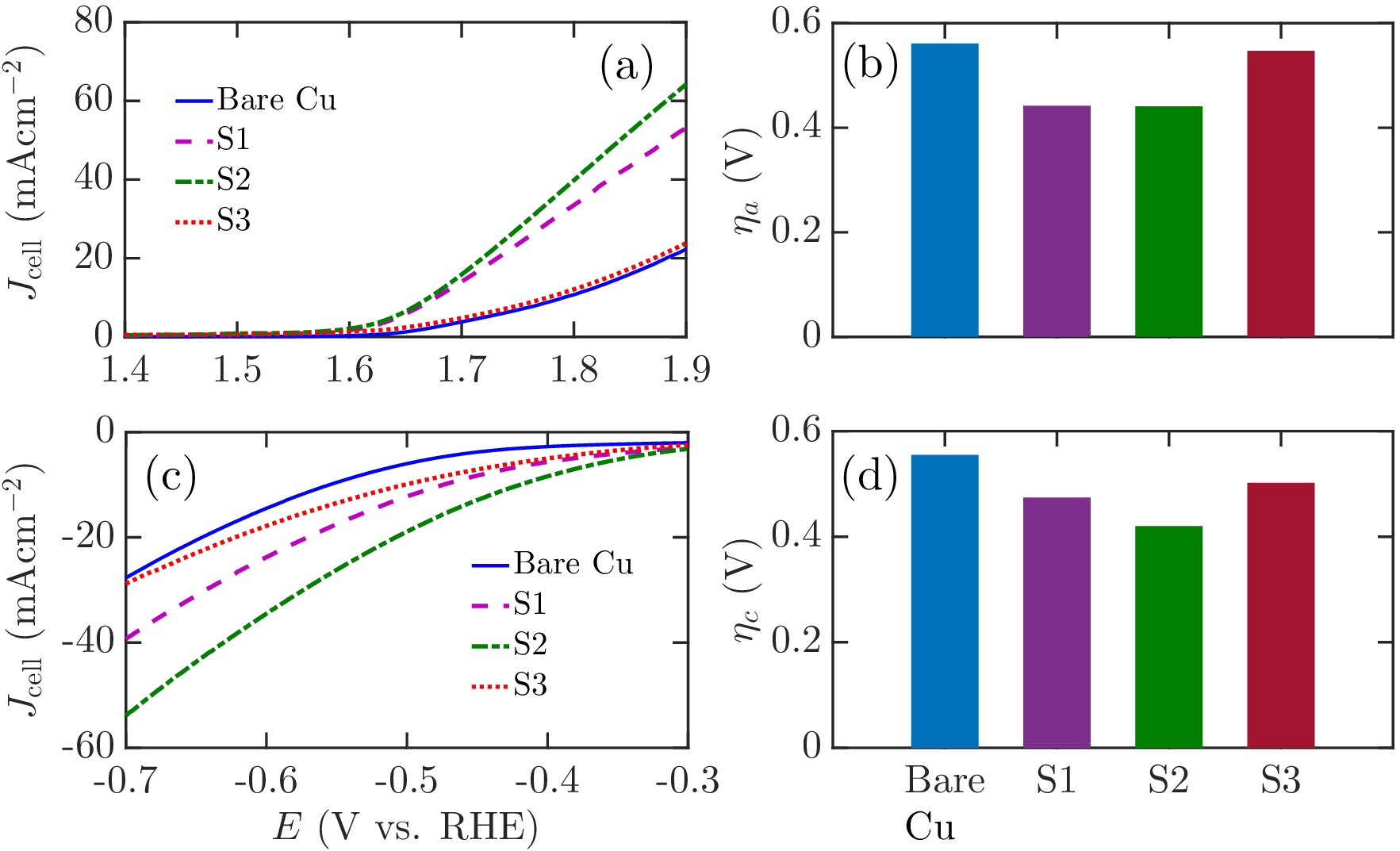}
    \caption{(a) Polarization curves (PC) and (b) overpotential losses ($\eta_a$) of bare and microstructured Cu electrodes for oxygen (O$_2$) evolution reaction (OER). (c) PC and (d) overpotential losses ($\eta_c$) of bare and microstructured Cu electrodes for the hydrogen (H$_2$) evolution reaction (HER). Both $\eta_a$ and $\eta_c$ were measured at $J_{\rm cell} = 10$ mAcm$^{-2}$. The concentration of the alkaline electrolyte (KOH) was 1 M.}
    \label{Fig 4}
\end{figure}

The electrochemical performance of the pristine and microstructured electrodes for both OER and HER was evaluated using PC, as illustrated in Figs.~\ref{Fig 4}(a) and \ref{Fig 4}(c). The curves show that S1 and S2 exhibit significant enhancement relative to the bare Cu electrode, whereas S3 indicates negligible improvement.  

Figures \ref{Fig 4}(b) and ~\ref{Fig 4}(d) illustrate the anodic overpotential, $\eta_a$, and the cathodic overpotential, $\eta_c$, for the bare and structured electrode samples, respectively. At $J_{\rm cell} =10$ mAcm$^{-2}$, S1 and S2 demonstrated $\eta_a$ of $0.442$ and $0.441$ V, respectively, resulting in a reduction in $\eta_a$ by $21.21\%$ and $21.39\%$ relative to bare Cu, respectively. This enhancement can be attributed to the increase in the number of active sites and better bubble detachment kinetics \cite{mamun2024enhancing, xie2026recent}. Since the number of microholes in S2 was greater than in S1, the number of active sites was higher, leading to enhanced OER and HER kinetics. Due to the confined space within the hole, a higher Laplacian pressure gradient develops, promoting bubble detachment \cite{li2024advances}. Additionally, the pressure gradient reduces bubble coalescence, thereby preventing surface blockage caused by bubble growth. Similarly, the reduction in $\eta_c$ for S1 and S2 relative to the pristine Cu electrode was $14.59\%$ and $24.32\%$, respectively. 

S3 had an $A_{\rm geo}$ of $4.6717$ mm$^2$ per unit cell according to Eq.~\eqref{eq:18}, which was $1.12$ and $1.36$ times larger than S1 and S2, respectively. Despite the largest $A_{\rm geo}$ per unit cell, S3 showed $\eta_a$ and $\eta_c$ of $0.547$ and $0.502$ V, respectively, indicating that the enhancement in electrocatalytic activities due to microstructuring is sub-optimal. S3 exhibits only $2.49\%$ and $9.55\%$ improvement in $\eta_a$ and $\eta_c$ relative to bare Cu, respectively. The larger $\eta_a$ and $\eta_c$ compared to S1 and S2 are due to slower kinetics of bubble detachment. The larger hole results in a lower Laplacian pressure gradient, which degrades detachment kinetics \cite{li2024advances}. Therefore, a greater bubble coalescence effect is achieved, resulting in greater surface blockage. S1 and S2 showed significant enhancements in OER and HER relative to bare Cu, which demonstrated their bifunctional electrocatalytic activity in alkaline water electrolysis.

\subsection{Electrocatalytic Activity and Reaction Kinetics}
\begin{figure}
    \centering
    \includegraphics[width =\linewidth]{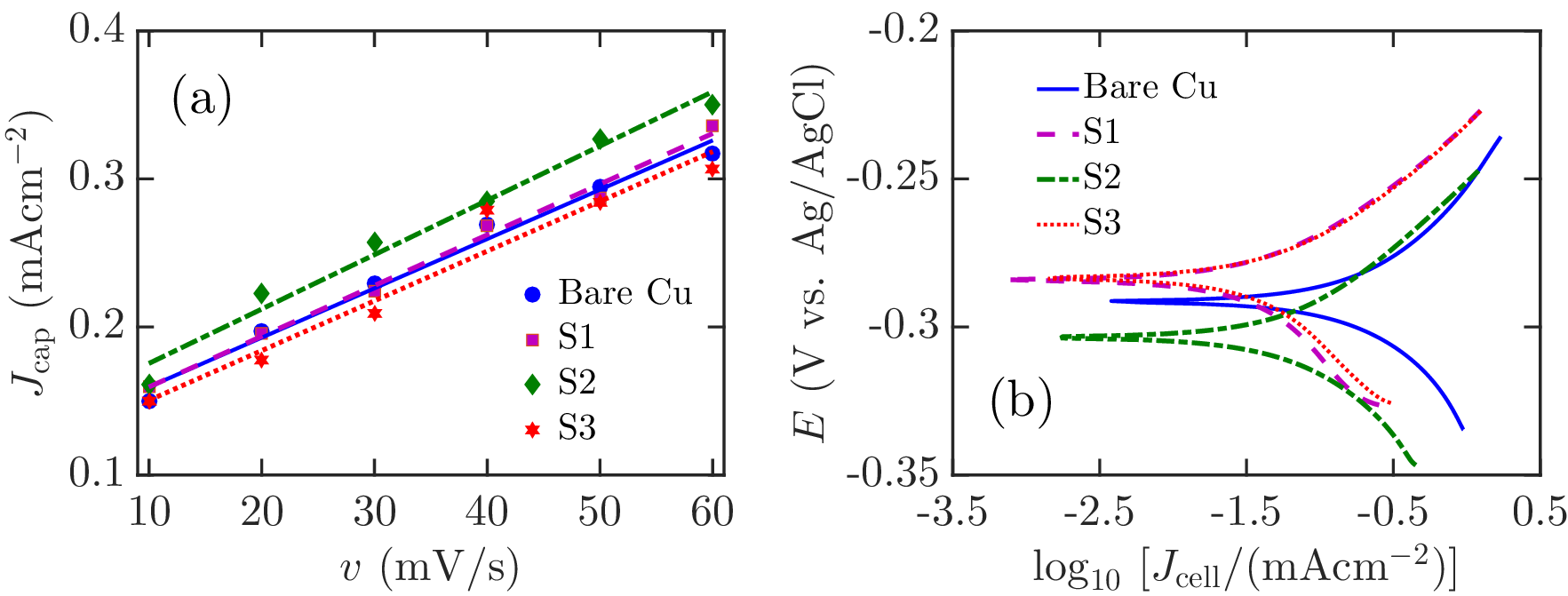}
    \caption{(a) The experimental data from cyclic voltammetry (CV) and the correlation line between capacitive current, $J_{\rm cap}$, and the scan rate, $v$. (b) The linear polarization resistance (LPR) analysis of the bare and microstructured Cu electrodes. The experiments were carried out in 1 M KOH solution at room temperature 25 $^ {\circ}$C.}
    \label{Fig 5}
\end{figure}
CV and LPR were performed to analyze the catalytic response and reaction rates of the bare and microstructured electrodes, which have been illustrated in Fig.~\ref{Fig 5}. A straight line was fitted between the capacitive current, $J_{\rm cap}$, and $v$, which is given as follows
\begin{equation}
    J_{\rm cap} = a_1v+a_2,
    \label{eq:25}
\end{equation}
where $a_1$ and $a_2$ are constants. The parameters, extracted from the experimental data, are shown in Table \ref{Table 3}.

The $E_{\rm A_{geo}}$ for S1 was $1.01\%$, while it had an $E_{\rm A_{EC}}$ of $3\%$. Additionally, despite a smaller $E_{\rm A_{geo}}$, S2 demonstrated improved catalytic performance. The $E_{\rm A_{EC}}$ of S2 was over $10\%$, which indicates a high number of active sites on the electrode surface. Although S3 had the highest $E_{\rm A_{geo}}$ of $4.10\%$ among all of its microstructured counterparts, its $E_{\rm A_{EC}}$ was $-2.40\%$, demonstrating that an increase in $A_{\rm geo}$ does not translate to a similar increase in $A_{\rm EC}$. While $A_{\rm geo}$ only depends on the geometry dimensions, $A_{\rm EC}$ is influenced by the surface roughness, wettability, electrolyte accessibility, and bubble accumulation \cite{ross2025impact}. 
\begin{longtable}{lllllll}
\caption{Electrocatalytic and kinetic parameters obtained from CV and LPR measurements. \label{Table 3}} \\
\hline \hline
Electrode & $C_{\rm dl}$ & $E_{\rm A_{\rm geo}}$ & $E_{\rm A_{\rm EC}}$ & $E_{\rm corr}$ & $b_a$ & $b_c$ \\
Sample & (mFcm$^{-2}$) & (\%) & (\%) & (V vs.~Ag/AgCl) & (mVdecade$^{-1}$) & (mVdecade$^{-1}$) \\
\hline
Bare Cu & 3.33 & - & - & $-0.29$ & 40.25 & 40.52 \\
S1 & 3.43 & 1.01\% & 3\% & $-0.28$ & 29.63 & 35.54 \\
S2 & 3.67 & 0.54\% & 10.21\% & $-0.3$ & 26.53 & 27.87 \\
S3 & 3.25 & 4.10\% & $-$2.40\% & $-0.28$ & 37.08 & 39.26 \\
\hline \hline
\end{longtable}
Among all the electrodes, S2 showed the highest $E_{\rm A_{EC}}$, suggesting an improved number of active sites, greater catalytic activity, higher charge transfer rates, and rapid bubble detachment kinetics. The LPR data, as shown in Table \ref{Table 3}, also demonstrated these enhancing effects. The $b_a$ and $b_c$ for S2 were the lowest among the electrodes, which were $26.53$ and $27.87$ mVdecade$^{-1}$, respectively, indicating faster electrochemical reaction kinetics and reduced activation energy barrier. This enhancement promotes higher rates of gas evolution reactions, including O$_2$ and H$_2$ evolution. Additionally, the lower $E_{\rm corr}$ relative to pristine Cu indicates a higher rate of electrochemical corrosion \cite{hussain2025physics}.     

However, S3 exhibited a negative $E_{\rm A_{EC}}$, which demonstrated its reduced catalytic activity compared to the bare electrode. Apart from this, $b_a$ and $b_c$ of S3 were comparable to those of bare Cu, indicating a small change in electrocatalytic activity, which can be attributed to poor bubble detachment kinetics. Due to sluggish detachment kinetics, the bubbles get trapped in the microholes, degrading surface catalytic activity. Therefore, the ions in the electrolyte cannot access the electrode surface, thereby decreasing its wettability. 

\subsection{Interfacial Charge Transfer}
\begin{figure}
    \centering
    \includegraphics[width =\linewidth]{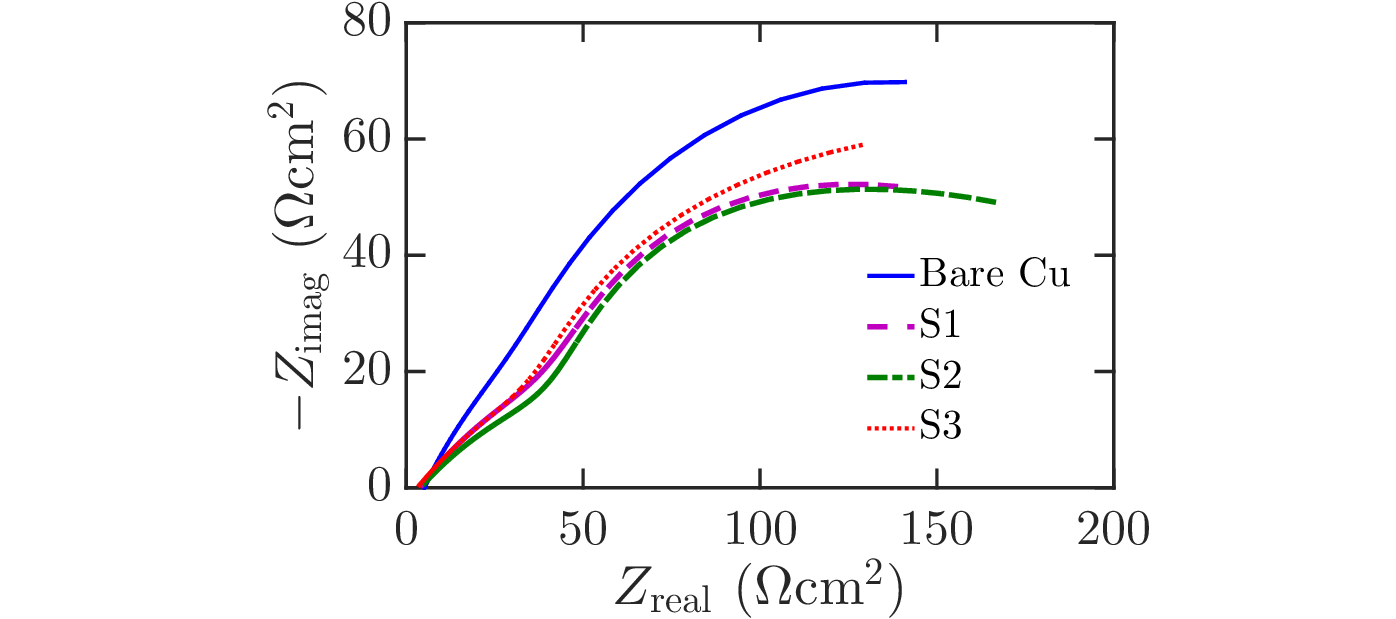}
    \caption{Nyquist plot of bare and microstructured Cu electrodes at the open-circuit potential (OCP) in 1 M KOH solution. The frequency range was from 0.01 Hz to 100 kHz, and the room temperature was 25 $^{\circ}$C.}
    \label{Fig 6}
\end{figure}

The EIS measurements were performed to gain insights into the behavior of the electrode surface and the electrode-electrolyte interface. The Nyquist plot is shown in Fig.~\ref{Fig 6}. A ladder-type RC network was fitted to the experimental data, as shown in Table \ref{Table 4} \cite{lazanas2023electrochemical, syam2023combination}. The solution resistance, $R_s$, of the bare and structured electrodes ranged between 3.11--5.07 $\Omega$cm$^2$ experimentally. The theoretical value of $R_s$ was $4.54$ $\Omega$cm$^2$, which closely aligned with the experimental results \cite{hodges2023critical}. 

The passivation layer resistance, $R_p$, and the passivation capacitance, $C_p$, characterize the passivation layer formed on the electrode surface and provide insight into its stability. A relatively higher $R_p$ indicates stronger resistance to electrochemical corrosion, which is desirable for the long-term stability of electrode materials \cite{hussain2025physics}. However, if $R_p$ is very high, the ions present in the electrolyte will face a higher resistance to diffuse towards the electrode surface. As a result, the electrode will suffer from poor electrolyte accessibility effects. Bare Cu had an $R_p$ of $140.26$ $\Omega$cm$^2$, which is the lowest among all the electrode samples, thereby demonstrating a high tendency for corrosion in the pristine metal. Apart from this, S1 and S2 have moderate $R_p$ of $154.07$ and $190.61$ $\Omega$cm$^2$, respectively, which is optimal for enhanced electrochemical performance. Meanwhile, S3 suffers from a thick passivation layer, which reduces carrier exchange between the interface and the bulk electrolyte.   
\begin{longtable}{p{0.8in} p{0.6in} p{0.6in} p{0.7in} p{0.4in} p{0.6in} p{0.75in}}
\caption{Extracted parameters of the electrochemical impedance spectroscopy (EIS) measurements. \label{Table 4}} \\
\hline \hline
Electrode & $R_s$ & $R_p$ & $C_p$ & $\tau_p$ & $R_{\rm ct}$ & CPE$_{\rm dl}$ \\
Sample & ($\Omega$cm$^2$) & ($\Omega$cm$^2$) & (mFcm$^{-2}$) & (s) & ($\Omega$cm$^2$) & ($\Omega$$^{-1}$cm$^{-2}$) \\
\hline
Bare Cu & $5.07$ & $140.26$ & $16.14$ & $2.26$ & $129.97$ & $17.03$ $s^{0.58}$ \\
S1 & $3.74$ & $154.07$ & $14.75$ & $2.27$ & $116.17$ & $11.39$ $s^{0.43}$ \\
S2 & $4.01$ & $190.61$ & $6.05$ & $1.15$ & $106.9$ & $7.87$ $s^{0.38}$ \\
S3 & $3.11$ & $226.77$ & $12.67$ & $2.87$ & $126.64$ & $14.09$ $s^{0.41}$ \\
\hline
\hline
\end{longtable}

The relaxation time of the passivation layer, $\tau_p$, is given by the following expression \cite{lazanas2023electrochemical}
\begin{equation}
    \tau_p = R_pC_p.
    \label{eq:26}
\end{equation}
Due to the transfer of ions and other charge carriers between the electrode-electrolyte interface and the bulk electrolyte, the interface undergoes charging and discharging. A lower $\tau_p$ indicates that the electrode surface responds rapidly to these phenomena. Among all the electrode samples, S2 had the lowest $\tau_p$ of $1.15$ s, indicating high responsivity to electrochemical reactions. However, S3 had the highest $\tau_p$, which was $2.87$ s, indicating a poor response to interfacial charge-transfer processes. Both bare Cu and S1 have nearly equal $\tau_p$. 

The most crucial parameter of the EIS measurements is $R_{\rm ct}$. It provides a qualitative insight into the charge transport phenomena at the electrode-electrolyte interface. Additionally, it indicates the surface reactivity and reaction kinetics \cite{rahman2025enhanced}. The $R_{\rm ct}$ of bare Cu, measured by our electrochemical measurements, was $129.97$ $\Omega$cm$^2$, which agrees well with the previously reported value of $135.6$ $\Omega$cm$^2$ \cite{syam2023combination}. The high $R_{\rm ct}$ exhibited by the bare Cu electrode shows a sluggish carrier transport mechanism. On the contrary, S2 had an $R_{\rm ct}$ of $106.9$ $\Omega$cm$^2$, smallest among the microstructured samples, signifying a rapid movement of charge carriers and promoting higher gas evolution rates. Additionally, S1 and S3 showed an $R_{\rm ct}$ of $116.17$ and $126.64$ $\Omega$cm$^2$, respectively. The slightly lower $R_{\rm ct}$ of S3 compared to bare Cu is consistent with the electrochemical measurements shown in Secs.~4.1 and 4.2.

\subsection{Model Validation}
The developed multiphysics CFD-electrochemical model was used to predict $\eta_a$ and $E_{\rm A_{EC}}$ for the microstructured electrodes S1, S2, and S3. The model was solved considering 2D, 2D (Curved), and 3D geometric structures, as shown in Fig.~\ref{Fig 2}. The theoretical and experimental results are presented in Table \ref{Table 5}. Here, we have considered the mean values of $D$ and $d$, presented in Table \ref{Table 2}.   
\begin{longtable}{p{0.7in} p{0.9in} p{0.8in} p{0.9in} p{0.85in} p{0.7in}}
\caption{Comparison between experimental and calculated values of $\eta_a$ and $E_{\rm A_{EC}}$ across 2D, 2D (Curved), and 3D geometric structures. \label{Table 5}} \\
\hline \hline
Electrode & Geometry & Parameter & Experimental & Calculated & Deviation \\
Sample & & & Value & Value & \\
\hline
\multirow{6}{*}{S1} & \multirow{2}{*}{2D} & $\eta_a$ (V) & $0.442$ & $0.488$ & $10.5\%$ \\
& & $E_{\rm A_{EC}}$ & $3.00\%$ & $6.14\%$ & \\
& \multirow{2}{*}{2D (Curved)} & $\eta_a$ (V) & $0.442$ & $0.489$ & $10.6\%$ \\
& & $E_{\rm A_{EC}}$ & $3.00\%$ & $4.43\%$ & \\
& \multirow{2}{*}{3D} & $\eta_a$ (V) & $0.442$ & $0.469$ & $6.3\%$ \\
& & $E_{\rm A_{EC}}$ & $3.00\%$ & $2.71\%$ & \\
\hline
\multirow{6}{*}{S2} & \multirow{2}{*}{2D} & $\eta_a$ (V) & $0.441$ & $0.485$ & $10.0\%$ \\
& & $E_{\rm A_{EC}}$ & $10.21\%$ & $5.04\%$ & \\
& \multirow{2}{*}{2D (Curved)} & $\eta_a$ (V) & $0.441$ & $0.486$ & $10.1\%$ \\
& & $E_{\rm A_{EC}}$ & $10.21\%$ & $3.84\%$ & \\
& \multirow{2}{*}{3D} & $\eta_a$ (V) & $0.441$ & $0.468$ & $6.0\%$ \\
& & $E_{\rm A_{EC}}$ & $10.21\%$ & $9.07\%$ & \\
\hline
\multirow{6}{*}{S3} & \multirow{2}{*}{2D} & $\eta_a$ (V) & $0.547$ & $0.492$ & $10.1\%$ \\
& & $E_{\rm A_{EC}}$ & $-2.40\%$ & $2.47\%$ & \\
& \multirow{2}{*}{2D (Curved)} & $\eta_a$ (V) & $0.547$ & $0.494$ & $9.8\%$ \\
& & $E_{\rm A_{EC}}$ & $-2.40\%$ & $-2.36\%$ & \\
& \multirow{2}{*}{3D} & $\eta_a$ (V) & $0.547$ & $0.472$ & $13.7\%$ \\
& & $E_{\rm A_{EC}}$ & $-2.40\%$ & $-0.39\%$ & \\
\hline
\hline
\end{longtable} 

The calculated and experimental values of $\eta_a$ for S1 were 0.488 and $0.442$ V, respectively, considering 2D geometry. The deviations between experimental and calculated values were 3.94\%, 10.5\%, and 10.6\% for 2D and 2D (Curved) structures, respectively. These deviations were further reduced to 6.3\% when a 3D structure was considered. Meanwhile, the $E_{\rm A_{EC}}$ of S1 was $2.71\%$ for the 3D structure, which provided excellent agreement with the experimental result. These analyses indicated that the geometrically representative simulation structures decreased the deviation between model predictions and experimental results.

Similar to S1, the 3D structure of S2 exhibited better agreement with experimental $\eta_a$, showing a variation of only $6.0\%$. Unlike S1, the 2D geometry provided a better match with the experimental results of $E_{\rm A_{EC}}$ compared to the 2D (Curved) geometry for S2. As shown in Fig.~\ref{Fig 3}(b), the interhole separation was different across horizontal and vertical directions, influencing bubble dynamics in the microstructured electrode. Since the 2D and 2D (Curved) models represent only a cross-section of the 3D model, they cannot capture 3D effects accurately, leading to greater variability with the experimental data than S1. However, the 3D structure, which was more geometrically and electrochemically representative, yielded an $E_{\rm A_{EC}}$ of $9.07\%$. The predicted value was closely aligned with the experimental $E_{\rm A_{EC}}$ of $10.21\%$, thereby demonstrating the accuracy of the model across variations in hole numbers.  

The microholes of S3 are, on average, $4.5$ and $5.6$ times larger than those of S1 and S2, respectively. Therefore, these microholes exhibit more pronounced variations in edge curvature, taper angle, and sidewall morphology than those of S1 and S2, due to the use of the SLM technique \cite{diver2004micro, mincuzzi2019novel}. The model predictions of $\eta_a$ and $E_{\rm A_{EC}}$ for S3, assuming a 2D structure, showed considerable variations from the experimental data. The calculated and experimental values of $\eta_a$ for this geometric structure were $0.492$ and $0.547$ V, respectively. The deviation between the experimental result and the model prediction was 10.10\%. Meanwhile, the calculated $E_{\rm A_{EC}}$ was positive, while the experimental $E_{\rm A_{EC}}$ was negative. Given the simplified rectangular sidewall morphology, the 2D simulation results had low predictive accuracy. 

Unlike those for S1 and S2, the 3D simulation results for S3 did not closely align with the experimental findings. The deviation for $\eta_a$ was 13.7\%, which demonstrated a notable difference between theory and experiment. However, the predicted $E_{\rm A_{EC}}$ was $-0.39\%$, which was closer to the experimentally measured value of $-2.40\%$. Therefore, the 3D structure was more reliable than the 2D structure for predicting electrochemical parameters because it provided a more realistic and higher-dimensional representation of the electrolyzer system. However, the 3D model did not account for fabrication-induced tapering and curvature variations, thereby deviating from the idealized cylindrical geometry and leading to larger discrepancies with the experimental results compared to S1 and S2. 

On the contrary, the 2D (Curved) geometry for S3 yielded an $E_{\rm A_{EC}}$ of $-2.36\%$ and a deviation in $\eta_a$ of 9.8\%, which provided the best match with the experimental data. The 2D (Curved) geometry captures fabrication-induced variations and non-uniform curvature effects more effectively than the idealized 3D model, thereby providing the best agreement with the experimental observations. 

Table \ref{Table 2} demonstrates that S3 had a broader distribution of $D$ than the other electrodes. Therefore, we solved the model again for the 2D (Curved) structure, considering $D = 430.8$ and $D = 634.2$ $\mu$m, the lowest and highest values of $D$ for S3, as shown in Table \ref{Table 2}. We noticed that for $D = 634.2$ $\mu$m, the deviation in $\eta_a$ decreased to 9.63\%. Despite the increase in $E_{\rm A_{EC}}$ to $-1.79\%$, the electrode having $D = 634.2$ $\mu$m provided a better overall match with the experimental results. This observation indicates that the electrochemical performance of S3 is disproportionately governed by its relatively larger microholes.

In addition to the simplified geometric structures, the variations between the experimental and calculated values of $\eta_a$ and $E_{\rm A_{EC}}$ arose because we assumed that the microstructured electrode was defect-free, devoid of edge defects, and had no local variations in grain orientations. However, in reality, it contains features such as edges, dislocations, and grain boundaries \cite{bishara2021understanding, chen2024work}. These imperfections can alter the metal's resistivity and the local work function, thereby affecting electron transfer at the microscale \cite{chen2024work}. As a result, the electrode's surface kinetics are affected, leading to a discrepancy between our model and the experimental results. 

\subsection{Engineering Implications}
Our coupled CFD-electrochemical model provides mechanistic insights into the enhancement in electrochemical performance due to microstructuring. All the results presented in this section were obtained using the 2D (Curved) geometry. 
\begin{figure}
    \centering
    \includegraphics[width =\linewidth]{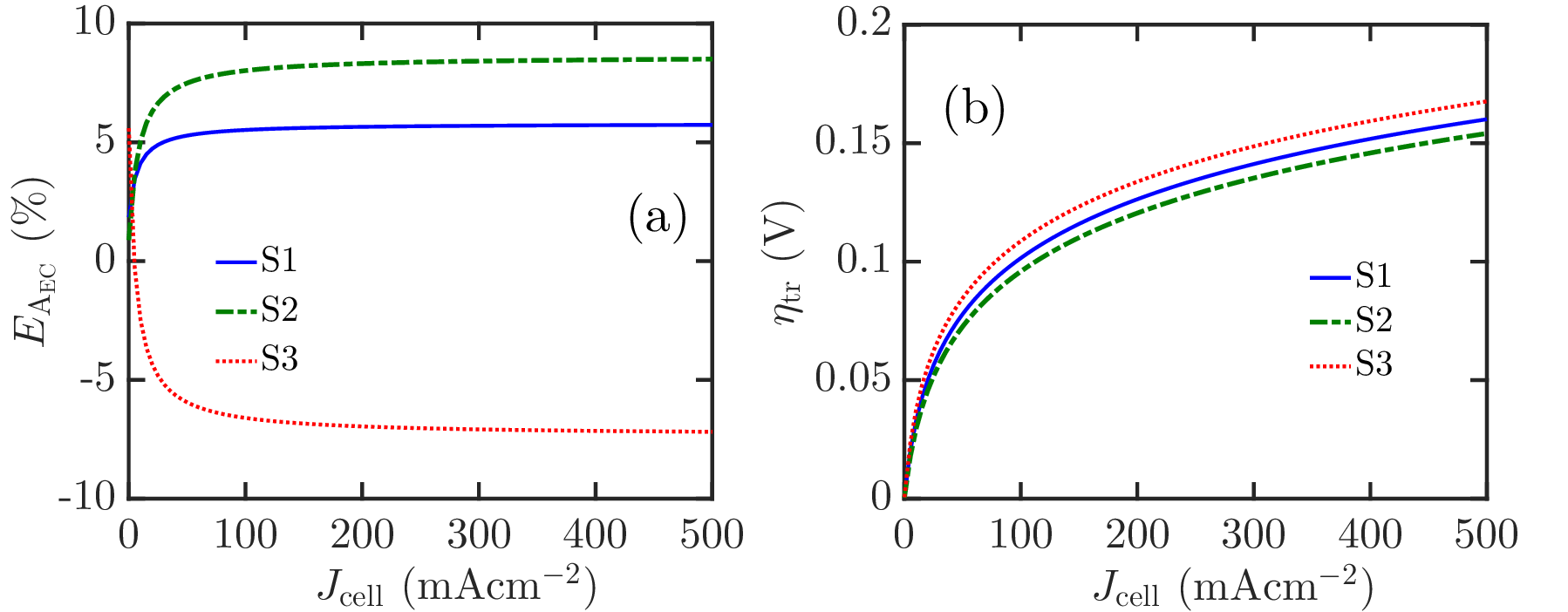}
    \caption{The dependence of (a) the enhancement in electrochemical surface area due to microstructuring ($E_{\rm A_{EC}}$) and (b) the transport overpotential loss ($\eta_{\rm tr}$) on the cell current ($J_{\rm cell}$). The concentration of the alkaline electrolyte is $1$ M, and the room temperature is 25 $^ {\circ}$C.}
    \label{Fig 7}
\end{figure}
Figure \ref{Fig 7} illustrates trends of $E_{\rm A_{EC}}$ and $\eta_{\rm tr}$ at varying $J_{\rm cell}$. Electrodes S1 and S2 demonstrated an increase in $E_{\rm A_{EC}}$ at low $J_{\rm cell}$, as shown in Fig.~\ref{Fig 7}(a), which indicated rapid bubble detachment kinetics, low bubble residence time, and weak coalescence effects. Additionally, the increased $E_{\rm A_{EC}}$ at low $J_{\rm cell}$ revealed that S1 and S2 were desirable for optimal gas evolution kinetics. However, as $J_{\rm cell}$ increases, bubble production becomes more prominent, leading to higher $\eta_{\rm tr}$, greater surface blockage, and sluggish detachment kinetics. Therefore, $E_{\rm A_{EC}}$ saturated at high $J_{\rm cell}$. S1 and S2 reached an $E_{\rm A_{EC}}$ of $5.734\%$ and $8.503\%$ at $J_{\rm cell} = 500$ mAcm$^{-2}$. On the contrary, S3 showed a decreasing trend in $E_{\rm A_{EC}}$ with increasing $J_{\rm cell}$, reaching an $E_{\rm A_{EC}}$ of $-7.183\%$ at $J_{\rm cell} = 500$ mAcm$^{-2}$. The decreasing trend of $E_{\rm A_{EC}}$ demonstrated that the microstructure offered a sluggish detachment mechanism and promoted bubble confinement. Additionally, Fig.~\ref{Fig 7}(b) showed the $\eta_{\rm tr}$ losses for S3 was higher compared to both S1 and S2. Therefore, S3 showed only a minor improvement over the pristine electrode in O$_2$ and H$_2$ evolution kinetics. 
\begin{figure}
    \centering
    \includegraphics[width =\linewidth]{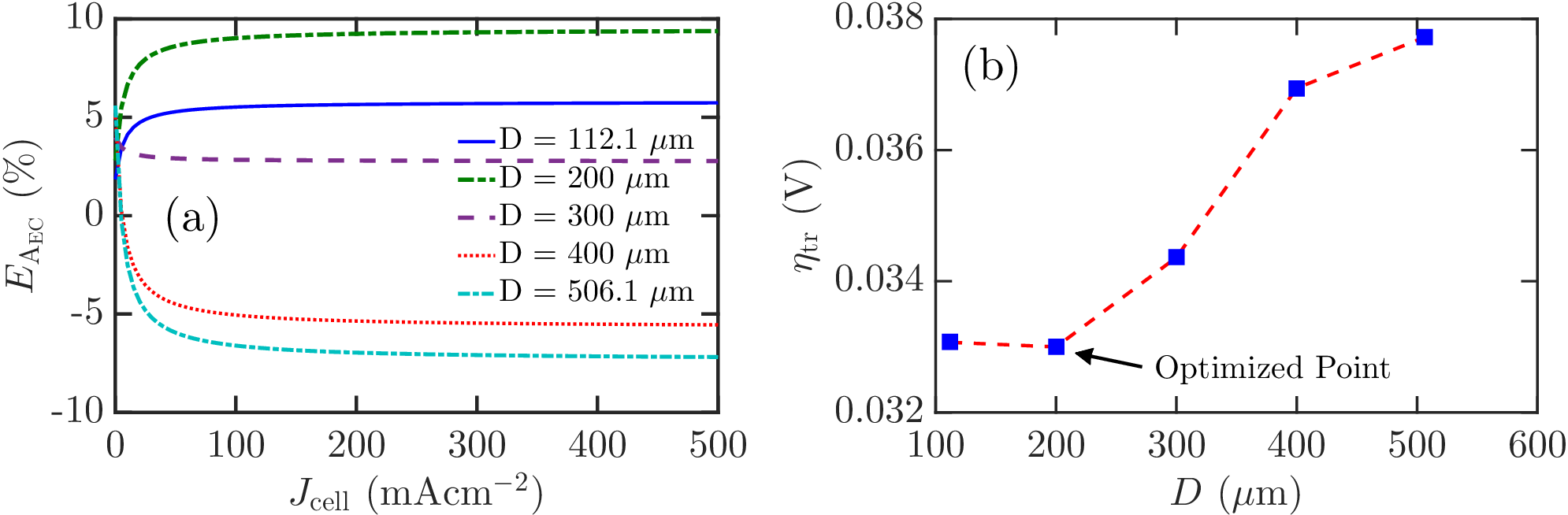}
    \caption{The dependence of (a) the enhancement in electrochemical surface area due to microstructuring ($E_{\rm A_{EC}}$) with varying microhole diameter ($D$). The depth of the microhole ($d$) was kept constant at $179.894$ $\mu$m. (b) Effect of $D$ on the transport overpotential loss ($\eta_{\rm tr}$) measured at $J_{\rm cell} = 10$ mAcm$^{-2}$. The concentration of the alkaline electrolyte was $1$ M, and the room temperature was 25 $^ {\rm o}$C.}
    \label{Fig 8}
\end{figure}

The difference in the trends shown by S1 and S3 in Fig.~\ref{Fig 7}(a) with varying $J_{\rm cell}$ indicates that there is an optimum $D$ for designing highly efficient microstructured electrodes. Therefore, we have varied $D$ from $112.1$ to $506.1$ $\mu$m, the mean $D$ of S1 and S3, respectively. This result has been illustrated in Fig.~\ref{Fig 8}(a). As $D$ increased from $112.1$ to $200$ $\mu$m, $E_{\rm A_{EC}}$ increased by $3.648\%$ at $J_{\rm cell} = 500$ mAcm$^{-2}$, indicating an enhancement in the number of activation sites on the electrode surface. However, at $D = 300$ $\mu$m, the trend shifted, and $E_{\rm A_{EC}}$ reduced to $2.779\%$, which was $\approx3\%$ below that of $D = 112.1$ $\mu$m at $J_{\rm cell} = 500$ mAcm$^{-2}$. This change in $E_{\rm A_{EC}}$ indicated the onset of poor detachment kinetics, leading to bubble-induced surface blockage and a degradation of $A_{\rm EC}$. As $D$ was increased further, $E_{\rm A_{EC}}$ decayed further, showing poor bubble detachment kinetics and leading to sub-optimal electrochemical performance compared to the pristine electrode. The $E_{\rm A_{EC}}$ for the microstructured electrodes having $D = 400$ $\mu$m and $D = 506.1$ $\mu$m were $-5.548\%$ and $-7.183\%$, respectively, exhibiting gas evolution kinetics similar to that of bare Cu. Figure \ref{Fig 8}(b) suggests that $\eta_{\rm tr}$ is minimum at $D=200$ $\mu$m. Therefore, $D=200$ $\mu$m is the optimal hole size for designing microstructured electrodes with $d = 179.894$ $\mu$m in a $4\times4$ pattern.

\begin{figure}
    \centering
    \includegraphics[width =\linewidth]{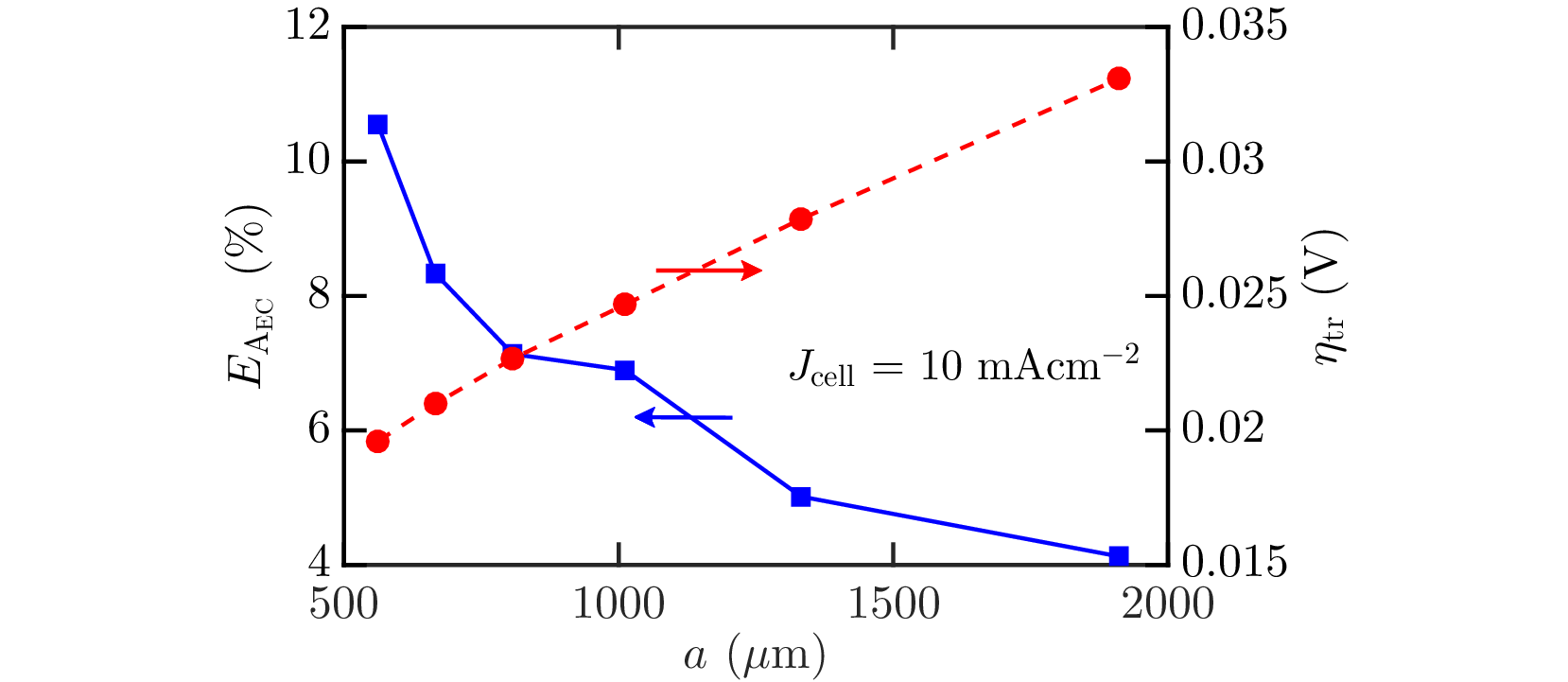}
    \caption{Effect of variation in the interhole separation ($a$) on the enhancement in electrochemical surface area due to microstructuring ($E_{\rm A_{EC}}$) and the transport overpotential loss ($\eta_{\rm tr}$) at $J_{\rm cell}$ = 10 mAcm$^{-2}$. The diameter ($D$) and the depth of the microhole ($d$) were $112.1$ and $179.894$ $\mu$m, respectively.}
    \label{Fig 9}
\end{figure}

In addition to $D$, we have varied $a$ to analyze the effect of interhole separation on the design of efficient bifunctional electrodes for water electrolysis. Figure \ref{Fig 9} shows the effect of $a$ on $E_{\rm A_{EC}}$ and $\eta_{\rm tr}$ at $J_{\rm cell} = 10$ mAcm$^{-2}$. The parameter $E_{\rm A_{EC}}$ was $10.54\%$ at $a=562$ $\mu$m. However, as $a$ increased from $562$ to $1910$ $\mu$m, $E_{\rm A_{EC}}$ exhibited a decreasing trend reaching $4.128\%$. The increase in $a$ can be attributed to a decrease in the density of microholes, which reduced surface reactivity. Meanwhile, $\eta_{\rm tr}$ demonstrated a systematic rise with increasing $a$, as shown in Fig.~\ref{Fig 9}, because of more pronounced bubble coalescence effects and sluggish bubble removal \cite{yang2025enhancing}.

\section{Conclusion}
Water electrolysis is one of the most promising technologies for renewable H$_2$ production. To overcome the electrocatalytic and kinetic limitations of conventional metal electrodes, microstructuring has emerged as an effective strategy for enhancing electrochemical performance. In this work, a multiphysics-based theoretical model integrating CFD and electrochemistry was developed to account for the effects of applied perturbation, surface confinement, surface coverage, and buoyancy. The model was used to analyze $E_{\rm A_{EC}}$ and $\eta_{\rm tr}$ for microstructured electrodes, and the results agreed well with the experimental findings. The model predicted an optimal $D$ of $200$ $\mu$m for a $4\times4$ hole configuration with a $d$ of $179.894$ $\mu$m. Furthermore, the model identified the dependence of electrocatalytic activity on $a$. These findings establish a physics-informed design framework for optimizing microstructured electrodes for industry-scale operations and provide valuable insights into the development of robust, cost-effective electrode architectures for sustainable H$_2$ generation.   

\section*{Data Availability}
All data of the paper are presented in the main text.

\section*{Author Declaration}
The authors have no conflicts of interest to disclose.
  
\small
\bibliographystyle{ieeetr}
\bibliography{references}
\end{document}